\hfuzz 2pt
\font\titlefont=cmbx10 scaled\magstep1
\magnification=\magstep1

\null
\vskip 1cm
\centerline{\titlefont ABOUT NON-POSITIVE EVOLUTIONS}
\medskip
\centerline{\titlefont IN OPEN SYSTEM DYNAMICS}
\vskip 2cm

\centerline{\bf F. Benatti}
\smallskip
\centerline{Dipartimento di Fisica Teorica, Universit\`a di Trieste}
\centerline{Strada Costiera 11, 34014 Trieste, Italy}
\centerline{and}
\centerline{Istituto Nazionale di Fisica Nucleare, Sezione di
Trieste}
\vskip .5cm

\centerline{\bf R. Floreanini}
\smallskip
\centerline{Istituto Nazionale di Fisica Nucleare, Sezione di
Trieste}
\centerline{Dipartimento di Fisica Teorica, Universit\`a di Trieste}
\centerline{Strada Costiera 11, 34014 Trieste, Italy}
\vskip .5cm

\centerline{\bf M. Piani}
\smallskip
\centerline{Dipartimento di Fisica Teorica, Universit\`a di Trieste}
\centerline{Strada Costiera 11, 34014 Trieste, Italy}
\centerline{and}
\centerline{Istituto Nazionale di Fisica Nucleare, Sezione di
Trieste}
\vskip 2cm

\centerline{\bf Abstract} 
\smallskip
\midinsert
\narrower\narrower\noindent
The long-time evolution of a system in interaction with an
external environment is usually described by a family
of linear maps $\gamma_t$,
generated by master equations of Block-Redfield type.
These maps are in general non-positive; a widely adopted cure 
for this physical inconsistency 
is to restrict the domain of definition of the dynamical maps
to those states for which $\gamma_t$ remains positive.
We show that this prescription has to be modified when two
systems are immersed in the same environment and evolve with
the factorized dynamics $\gamma_t\otimes\gamma_t$ starting from an 
entangled initial state.
\endinsert
\bigskip
\vfill\eject

\noindent
{\bf 1. INTRODUCTION}
\medskip

The dynamics of systems immersed in large, external environments
can be described in terms of master equations: they generate
the finite time-evolution for the reduced density matrix,
obtained by tracing over the environment degrees of freedom.[1-9]
Their explicit form is in general rather complex, involving
non-linearities and memory effects. Nevertheless, when the 
coupling between subsystems and environment is sufficiently weak
and for times much longer than the characteristic correlation-time
in the environment, suitable limiting master equations in Markovian 
form can be derived.

These derivations are often based on ad hoc approximations,
lacking mathematical rigor, while the final result is justified 
on the basis of physical considerations. Despite these heuristic 
treatments, Markovian master equations have been applied to model 
various effects in open system dynamics, ranging from quantum optics
to quantum chemistry.

It has been pointed out long ago that the heuristic derivations of the
Markovian limit of master equations could lead in general to physical
inconsistencies.[10] In particular, the resulting finite time-evolution
described by such equations would not in general preserve the positivity
of the reduced density matrix, with some remarkable exceptions,
based on rigorous mathematical treatments.[1-4, 11-19]

Although acknowledged in most subsequent literature on the subject,
these inconsistencies were either dismissed as irrelevant for all
practical purposes [20] or cured by adopting further ad hoc prescriptions.[21-23]
In the latter case, the general attitude is to restrict the action of the
dynamical maps generated by the non-positive master equations to a subset
of all possible initial reduced density matrices, those for which 
the time-evolution remains positive. This is equivalent to a suitable
selection of the initial conditions for the starting state of the subsystem,
a procedure sometimes referred to as ``slippage of the initial conditions''.
On physical grounds, this effect is viewed as the consequence of the
short-time correlations in the environment, that have not been
properly taken into account in the derivation of the Markovian limit
of the original master equation.%
\footnote{$^\dagger$}{Let us point out that, instead of restricting the
possible initial states, one can alternatively ``smooth'' the initial
conditions on which the non-positive dynamical map acts [24];
the resulting effective map turns out to be positive.
Unless it results also completely positive, unphysical effects of the kind
discussed below would affect this  case as well.}

In the following, we shall reexamine this widely used prescription to
cure possible inconsistencies produced by non-positive,
Markovian master equations, and point out further potential problems
of this approach. We shall deal with two identical, non-interacting
subsystems immersed in a same environment, both evolving, in the
Markovian limit, with the same non-positive master equation.
We shall explicitly show that redefining the initial conditions
to make positive the single system time-evolution is not enough 
to cure all possible inconsistencies of the two-system dynamics.
These show up when the two-system state that emerges after the
transient due to the short-time correlations in the environment
is entangled; therefore, in order to have a physically acceptable
time evolution for the two subsystems when entanglement is the
most likely consequence of the initial transient phase, the above 
mentioned procedure of restricting initial conditions should take 
into account also correlated states.

On the other hand, let us notice that maximal entanglement 
can be produced without any transient. A particularly
interesting example is that of two neutral kaons that are
produced, via the weak interaction, 
as decay products of a spin-one $\Phi$ resonance:
due to angular momentum conservation, the two spin-zero kaons fly
apart back to back, in a state that resembles that of the singlet
for two spin 1/2 particles.[25]

Using standard techniques, in the next Section we shall derive
the Markovian limit of the master equation describing two 
two-level systems in interaction with a stochastic environment.
After waiting for the correlations in the environment to die out,
the resulting finite time-evolution $\Gamma_t$ turns out to be
describable in terms of a factorized dynamics: 
$\Gamma_t=\gamma_t\otimes\gamma_t$, where $\gamma_t$ represent
a single open system dynamics, in general non-positive.
In Section 3 we shall then apply the derived time-evolution $\Gamma_t$ 
to a maximally entangled, pure initial state, and show that ``negative
probabilities'' may arise even though the dynamics
$\gamma_t$ remains positive on the constituent single-system states.
The case of partially entangled states and that of mixed entangled states
is discussed in Section 4. The concluding Section 5 contains our 
final considerations.

\vskip 1cm

\noindent
{\bf 2. MARKOVIAN MASTER EQUATION}
\medskip

The physical model we shall study is formed by
two, non-interacting, two-level systems immersed in the
same, external environment. The Markovian limit
of their subdynamics will be derived using the same
techniques and approximations widely adopted in
analyzing single system time-evolutions.[5-9]

For sake of definiteness, the action of the environment
on the two subsystems will be assumed to be mediated by a weak 
time-dependent stochastic field, coupled to their spin-like
degrees of freedom.[26-28] This choice is of sufficient generality for
the considerations that follow. Let us point out that 
this model can describe real physical situations, 
like the ones occurring in interferometric set-ups, involving the
propagation of neutrons in random magnetic fields [29-36] or photons in
random optical media.[5-8, 37, 38] 
Moreover, it has been used to study dissipative effects in correlated neutral mesons under the action of weak, stochastic gravitational fields.[39-42]

Without loss of generality, the total system hamiltonian can be taken to be:[36]
$$
H=H_0^{(1)} + H_0^{(2)} + H_I^{(1)} + H_I^{(2)}\ ,
\eqno(2.1)
$$
$$
H_0^{(A)}={\omega_0\over 2}\,\sigma_1^{(A)}\ ,\qquad
H_I^{(A)}=\sum_{i=1}^3 V_i^{(A)}(t)\,\sigma_i^{(A)}\ ,\qquad
A=1,2\ ,
\eqno(2.2)
$$
where $\sigma_i^{(1)}=\sigma_i\otimes{\bf 1}$, 
$\sigma_i^{(2)}={\bf 1}\otimes\sigma_i$, are the two system spin operators,
represented by the Pauli matrices $\sigma_i$, $i=1,2,3$, while
${\bf V}^{(A)}(t)=(V_1^{(A)}(t),V_2^{(A)}(t),V_3^{(A)}(t))$, $A=1,2$,
are the stochastic, Gaussian field variables, independently coupled
to the spin degrees of freedom of the two systems.
For simplicity, we assume ${\bf V}^{(A)}(t)$ to have  zero mean, 
$\langle{\bf V}^{(A)}(t)\rangle=0$, and 
stationary, real, positive-definite covariance matrix 
$[W_{ij}^{(AB)}(t)]$
with entries 
$$
W_{ij}^{(AB)}(t-s)=\langle V_i^{(A)}(t)V_j^{(B)}(s)\rangle=
\big(W_{ij}^{(AB)}\big)^*(t-s)=W_{ji}^{(BA)}(s-t)\ .
\eqno(2.3)
$$

For completeness, let us point out that the hamiltonian (2.1) can be
equivalently interpreted as describing two subsystems in interaction
with two independent baths of identical physical characteristics;
for defineteness, in the following we find more convenient to refer
to the single bath picture.

Being coupled to a stochastic field, the complete $4\times 4$ 
spin density matrix $R(t)$ is also stochastic;
an effective, ``reduced'', spin density matrix $\rho(t)$
is obtained by averaging over the noise: $\rho(t):=\langle R(t)\rangle$.
At the initial time $t=0$ we may suppose spin and noise to
decouple, so that:
$\rho\equiv\langle R(0)\rangle=R(0)$.

The dynamical equation for $\rho(t)$ can be obtained in a standard way
from the usual Liouville-von Neumann equation for $R(t)$, through
the intermediate use of the interaction picture.
The resulting master equation contains an infinite series of terms.
As usually done in the case of a single system subdynamics, a simplified, more manageable expression for it can be derived by means of
physical considerations.[5-9, 26]

By hypothesis, the action of the external stochastic field on the
two subsystems is weak; within this ``weak coupling limit'' assumption, 
one can then focus
on the dominant first term in the expansion of the general master equation, neglecting higher order contributions. One explicitly finds:
$$
\eqalignno{
&\partial_t\rho(t)=-i\Bigl[H_0^{(1)}+H_0^{(2)},\ \rho(t)\Bigr]\,-\,
\sum_{A,B=1}^2\sum_{i,j=1}^3C_{ij}^{(AB)}(t)\Bigl[\sigma_i^{(A)},
\Bigl[\sigma_j^{(B)},\ \rho(t)\Bigr]\Bigr]\ ,
&(2.4a)\cr
&C_{ij}^{(AB)}(t)=\sum_{k=1}^3\int_0^t{\rm d}s\, W_{ik}^{(AB)}(s)
\,U_{k j}(s)\ , &(2.4b)}
$$
where 
$$
U_{ij}(t)=\pmatrix{1 & 0 & 0\cr
0 & \cos\omega_0 t & \sin\omega_0 t\cr
0 & -\sin\omega_0 t & \cos\omega_0 t\cr}
\eqno(2.5)
$$
is the orthogonal matrix $[U_{ij}(t)]$ that
represents the rotations of the Pauli matrices due to the action of
the free Hamiltonian:
$
e^{-itH_0^{(A)}}\sigma_i^{(A)}e^{itH_0^{(A)}}=
\sum_{j=1}^3U_{ij}(t)\, \sigma_j^{(A)}$.

Further, by the same physical arguments, the memory effects 
in $(2.4)$ should not be physically relevant: within the abovementioned
hypothesis, the use of the Markovian limit is therefore justified;
in practice, this can be implemented by extending to infinity 
the upper limit of integration in $(2.4b)$.
More precisely, for situations amenable to a rigorous mathematical treatment, 
one can show that a linear, local in time
subdynamics is the general result of a limiting procedure in which the coupling
constant $\xi$ between system and external environment, and the ratio
$\tau/T$ between the typical time scale of the system and the decay time
of the correlations in the environment, become small.[2-4]
The quantities $\xi$ and $\tau/T$ regulate both the weak coupling limit
and the Markovian approximation.

In order to keep the discussion in the subsequent sections as simple as possible, we shall make some further simplifying assumptions on the environment correlations (2.3). We first assume that the external stochastic field be oriented along the third direction, ${\bf V}^{(A)}(t)=(0,0,V_3^{(A)}(t))$, with exponentially
suppressed correlation functions:
$$
\eqalignno{
&\langle V_3^{(1)}(t)\, V_3^{(1)}(s)\rangle=
\langle V_3^{(2)}(t)\, V_3^{(2)}(s)\rangle=g^2\, e^{-\mu|t-s|}\ , &(2.6a)\cr
&\langle V_3^{(1)}(t)\, V_3^{(2)}(s)\rangle=f^2\, e^{-\nu|t-s|}\ . &(2.6b)}
$$
Furthermore, we make the physically sensible hypothesis that the non-diagonal,
off-site correlations $W^{(12)}_{33}$ be subdominant with respect to
the diagonal, on-site ones, $W_{33}^{(AA)}$; in practice, this can be achieved
by assuming a hierarchy in the strength $(f^2\ll g^2)$
and decay constants $(\mu\ll\nu)$ of the two types of correlations.
In this way, the interaction between the two subsystems induced
by the coupling with the environment becomes negligible.
(The general case is briefly treated in the Appendix, where more details on the derivation of Eq.(2.7) below can also be found.) 
Then, to lowest order, the finite time-evolution
for the density matrix, $\rho(0)\mapsto\rho(t)\equiv\Gamma_t[\rho(0)]$, 
assumes a factorized form,
 $\Gamma_t=\gamma_t\otimes\gamma_t$, being generated by the following
Markovian master equation:
$$
\partial_t\rho(t)=\big(L\otimes{\bf 1}+{\bf 1}\otimes L\big)[\rho(t)]\ ;
\eqno(2.7)
$$
the linear operator $L[\cdot]\equiv L_0[\cdot]+L_1[\cdot]$, the generator of $\gamma_t$, acts on $2\times2$ density matrices $\eta$, 
and its explicit form is as follows:
$$
\eqalignno{
&L_0[\eta]=-i\big[H_0,\eta\big]\ ,\qquad\quad H_0=\omega\, \sigma_1\ , &(2.8a)\cr
&L_1[\eta]=\alpha\big(\sigma_3\,\eta\,\sigma_3-\eta\big)
-\beta\big(\sigma_2\,\eta\,\sigma_3 + \sigma_3\,\eta\,\sigma_2\big)\ , &(2.8b)}
$$
with
$$
\alpha={2 g^2\mu\over\omega_0^2+\mu^2}\ ,\qquad
\beta={ g^2\omega_0\over\omega_0^2+\mu^2}\ ,\qquad
\omega={\omega_0\over2}+\beta\ .
\eqno(2.8c)
$$
In this way, each system evolves independently, with the dynamics
generated by%
\footnote{$^\dagger$}{It is interesting to notice that essentially the same master equation (2.8), (2.9) is also the result of the Markovian approximation of a stochastic dynamical evolution based on the `` quantum state diffusion'' approach.[43, 44]}
$$
\partial_t\eta(t)=L[\eta(t)]\ .
\eqno(2.9)
$$
This equation is of the Bloch-Redfield type [5-9] and as such it is known
not to be positive.%
\footnote{$^\ddagger$}{Indeed, for sufficiently small, but positive
times, the evolution equation (2.9) will map
the initial state $\eta(0)=\left(\matrix{ 1 & 0\cr 0 & 0\cr}\right)$
into a non-positive matrix $\eta(t)$.}
As already mentioned in the Introduction, to cure this pathology an ad hoc
prescription has been proposed, widely adopted in the literature:
restrict the possible initial states $\eta(0)$ to those for which
$\eta(t)\equiv\gamma_t[\eta(0)]$, $t>0$, as generated by (2.9), is still
a state. As we shall see in the next section, this requirement is in general
not enough to guarantee the consistency of (2.7).

\vfill\eject

\noindent
{\bf 3. MAXIMALLY ENTANGLED STATES}
\medskip

As shown in the previous section, the dynamics of two non-interacting systems immersed in the same bath takes a factorized form, 
$\Gamma_t=\gamma_t\otimes\gamma_t$, at least for times
much longer than the characteristic correlation times in the environment.
However, the initial state $\rho(0)$ of the compound system, on which $\Gamma_t$
acts, need not be in factorized form: due to the short-time interaction with the environment, the subsystems can emerge from the transient in an
entangled state.

To avoid inconsistencies with the single-system dynamics $\gamma_t$,
we shall adopt the previously mentioned prescription of restricting its action
to those states for which positivity is guaranteed, for any $t\geq0$.
For the compound system under study, this amounts to require that the partial traces $\eta^{(1)}={\rm Tr}_2[\rho(0)]$ and $\eta^{(2)}={\rm Tr}_1[\rho(0)]$
over the degrees of freedom of the second, respectively the first,
subsystem be admissible states for the map $\gamma_t$.

Let us decompose a generic $2\times2$ density matrix $\eta$ along the Pauli matrices and the identity $\sigma_0$:
$\eta=\sum_{\mu=0}^3\eta^\mu\sigma_\mu$, with $\eta^0=1/2$ and $\eta^i$ real;
its time-evolution, generated by the equation (2.8)--(2.9),
is then given by
$\eta(t)\equiv\gamma_t[\eta]=\sum_{\mu=0}^3\eta^\mu(t)\sigma_\mu$,
with components:
$$
\eqalignno{
&\eta^0(t)={1\over2}\ , &(3.1a)\cr
&\eta^1(t)=e^{-2\alpha t}\eta^1\ ,&(3.1b)\cr
&\eta^2(t)=e^{-\alpha t}\Bigg[\bigg(\cos2\Omega t-{\alpha\over2\Omega}
\sin2\Omega t\bigg)\eta^2 -{\omega+\beta\over\Omega}\sin2\Omega t\ \eta^3
\Bigg]\ ,&(3.1c)\cr
&\eta^3(t)=e^{-\alpha t}\Bigg[\bigg(\cos2\Omega t+{\alpha\over2\Omega}
\sin2\Omega t\bigg)\eta^3 +{\omega-\beta\over\Omega}\sin2\Omega t\ \eta^2
\Bigg]\ ,&(3.1d)\cr}
$$
with $\Omega=\sqrt{\omega^2-\beta^2-\alpha^2/4}$, and $\eta^1$, $\eta^2$, 
$\eta^3$ the initial density matrix components.

With the eigenstates 
$\eta_\pm\equiv |\pm\rangle\langle\pm|=(\sigma_0\pm\sigma_1)/2$
of the free systems hamiltonian in $(2.8a)$,
$H_0 |\pm\rangle=\pm\omega |\pm\rangle$, one can build the maximally
entangled state:
$$
|\psi\rangle={1\over\sqrt{2}}\Big(|+\rangle\otimes|-\rangle
- |-\rangle\otimes|+\rangle\Big)\ .
\eqno(3.2)
$$
Let us assume that after the transient the compound system be in such a state,
so that the initial density matrix, on which the total Markovian dynamics
$\Gamma_t=\gamma_t\otimes\gamma_t$ acts, is given by:
$$
\rho(0)\equiv |\psi\rangle\langle\psi|=
{1\over2}\Big(\eta_+\otimes\eta_- + \eta_-\otimes\eta_+
-\eta_{+-}\otimes\eta_{-+} - \eta_{-+}\otimes\eta_{+-}\Big)\ ,
\eqno(3.3)
$$
with $\eta_{\pm\mp}=(\sigma_3\pm i\sigma_2)/2$. 
The two partial traces 
$\eta^{(1)}={\rm Tr}_2[\rho(0)]$ and $\eta^{(2)}={\rm Tr}_1[\rho(0)]$,
being equal to $\sigma_0/2$, are left invariant by the dynamics (3.1), and therefore represent admissible states for the evolution $\gamma_t$.

It is a matter of a simple computation to apply the evolution given in (3.1) to the four $2\times2$ matrices $\eta_+$, $\eta_-$, $\eta_{+-}$, $\eta_{-+}$ and therefore obtain
the explicit expression for the evolved $4\times4$ matrix
$\rho(t)=\Gamma_t[\rho(0)]\equiv \gamma_t\otimes\gamma_t[\rho(0)]$.
In the basis for which the Pauli matrices assume the standard form,
$\sigma_1=\left(\matrix{0 & 1\cr 1 & 0}\right)$,
$\sigma_2=\left(\matrix{0 & -i\cr i & 0}\right)$,
$\sigma_3=\left(\matrix{1 & 0\cr 0 & -1}\right)$,
one explicitly gets:
$$
\rho(t)={1\over4}\left(\matrix{A_-(t) & C(t)  & C(t)  & B_+(t)\cr
                               -C(t)  & A_+(t) & B_-(t) & -C(t)  \cr
                               -C(t)  & B_-(t) & A_+(t) & -C(t)  \cr
                               B_+(t) & C(t)  & C(t)  & A_-(t)\cr}\right)\ ,
\eqno(3.4)
$$
where
$$
\eqalignno{
&A_\pm(t)=1\pm e^{-2\alpha t}
\Bigg[\bigg(\cos2\Omega t+{\alpha\over2\Omega}
\sin2\Omega t\bigg)^2 +\bigg({\omega-\beta\over\Omega}\bigg)^2
\sin^2 2\Omega t\Bigg]\ , &(3.5a)\cr
&B_\pm(t)=-e^{-4\alpha t}\pm e^{-2\alpha t}
\Bigg[\bigg(\cos2\Omega t-{\alpha\over2\Omega}
\sin2\Omega t\bigg)^2 +\bigg({\omega+\beta\over\Omega}\bigg)^2
\sin^2 2\Omega t\Bigg]\ , &(3.5b)\cr
&C(t)=ie^{-2\alpha t}\,\sin 2\Omega t\Bigg[
{2\beta\over\Omega}\cos 2\Omega t
-{\alpha\omega\over\Omega^2} \sin 2\Omega t\Bigg]\ . &(3.5c)}
$$

The matrix $\rho(t)$ in (3.4) should represent the state of the compound system
at time $t$, having been originally prepared in the initial entangled state
(3.3). The matrix (3.4) should then be positive. However, one can easily check that one of its eigenvalues can become negative,
precisely that corresponding to the eigenvector $(1,0,0,-1)$.%
\footnote{$^\dagger$}{Using the definitions in (3.5),
the four eigenvalues can be explicitly written as:
$(A_+ - B_-)/4$, $(A_- - B_+)/4$, 
$\big\{(A_+ + A_- + B_+ + B_-)\pm \big[(A_+ - A_- + B_- - B_+)^2 -16 C^2
\big]^{1/2}\big\}/8$. The corresponding eigenvectors turn out to be
time dependent; their explicit expressions are involved and not
particularly inspiring.}
Indeed, from its expression:
$$
\lambda(t)={1\over4}\Bigg\{1+e^{-4\alpha t}-2e^{-2\alpha t}\bigg[
\cos^2 2\Omega t+{2\omega^2-\Omega^2\over\Omega^2}\sin^2 2\Omega t\bigg]
\Bigg\}\ ,
\eqno(3.6)
$$
one checks that $\lambda(0)=\dot\lambda(0)=\,0$, while
$\ddot\lambda(0)=-8\beta^2$, so that $\lambda(t)$ starts assuming negative
values as soon as $t$ becomes nonzero.

In order for the map $\Gamma_t=\gamma_t\otimes\gamma_t$ to
produce a physically acceptable dynamics, states like (3.3) must therefore 
be excluded from its domain of definition. Entanglement is
crucial in revealing this physical inconsistency; indeed,
on factorized states $\eta^{(1)}\otimes\eta^{(2)}$, with
$\eta^{(1)}$, $\eta^{(2)}$ admissible starting density matrices for
$\gamma_t$, the dynamics $\Gamma_t=\gamma_t\otimes\gamma_t$
remains positive.

The negativity of $\lambda(t)$ will not last for ever:
due to the damping factors, the expression in (3.6) becomes
positive after a certain time, and actually asymptotically
tends to $1/4$, as the remaining three eigenvalues of (3.4).
This is a consequence of the dynamics generated by (2.7)
for which the von Neumann entropy, $S[\rho]=-\rho\ln\rho$,
always increases (as already observed, 
the map $\gamma_t$, hence $\Gamma_t$, is
unital, $\gamma_t[\sigma_0]=\sigma_0$); therefore, any initial state
$\rho(0)$ of the compound system is asymptotically driven
for long times to the maximally disordered state
$\rho=\sigma_0\otimes\sigma_0/4$. 

In general, master equations of the form (2.9), $(2.8b)$ may involve
parameters $\alpha$ and $\beta$, not as in $(2.8c)$, but totally
independent. In such cases, contrary to $(2.8c)$, $\alpha$ can
become vanishingly small, without conflicting with the
Markovian hypothesis. Consequently,
the eigenvalue $\lambda(t)$ becomes 
a periodic function of time and assumes
negative values even for arbitrary large times.

\vskip 1cm

\noindent
{\bf 4. PARTIALLY AND MIXED ENTANGLED STATES}
\medskip

As already observed, it is the entanglement of the initial state $\rho(0)$ of the otherwise independent two subsystems that allows revealing
the unphysical effect of production of ``negative probability'' by the
dynamics $\Gamma_t$. The magnitude of the phenomenon is directly connected
to the amount of entanglement that the initial state $\rho(0)$ contains.

This can be easily shown by taking the partially entangled state
$$
|\psi_\theta\rangle=\cos\theta\, |+\rangle\otimes|-\rangle
- \sin\theta\, |-\rangle\otimes|+\rangle\ ,
\eqno(4.1)
$$
as starting state, instead of the maximally entangled one in (3.2).
The evolution in time of the corresponding density matrix
$\rho_\theta(0)=|\psi_\theta\rangle\langle\psi_\theta|$ 
can be easily obtained as before using 
the explicit expressions in (3.1).%
\footnote{$^\dagger$}{Notice that as before the partial traces
$\eta^{(1)}_\theta={\rm Tr}_2[\rho_\theta(0)]=
\sigma_0/2+\cos2\theta\,\sigma_1/4$ and 
$\eta^{(2)}_\theta={\rm Tr}_1[\rho_\theta(0)]=
\sigma_0/2-\cos2\theta\,\sigma_1/4$ are perfectly admissible states
of the dynamics $\gamma_t$.}
One finds that also in this case the eigenvalue $\lambda_\theta(t)$ of 
$\rho_\theta(t)=\Gamma_t[\rho_\theta(0)]$ corresponding to the eigenvector $(1,0,0,-1)$
can assume negative values;
in fact, $\lambda_\theta(0)=\dot\lambda_\theta(0)=\,0$, while
$\ddot\lambda_\theta(0)=2\big(\alpha^2 \cos^2 2\theta
-4\beta^2\sin^2 2\theta\big)$, which is negative provided: 
$\tan^2 2\theta\geq \alpha^2/4\beta^2$.

In other terms, once the evolution $\gamma_t$ is given, and therefore the parameters $\alpha$ and $\beta$ of the corresponding master equation
are fixed, the time evolution $\Gamma_t=\gamma_t\otimes\gamma_t$ of the
compound system becomes physically inconsistent on initial states that
possess a sufficiently high degree of entanglement.
Therefore, in order for $\Gamma_t$ to be an acceptable Markovian evolution,
one has to further restrict its domain of definition, in order to exclude
also those partially entangled states.

The discussion can be extended to entangled mixed initial states,
like the Werner states:[45]
$$
\rho_W=p\,\rho + {1-p\over4}\, \sigma_0\otimes\sigma_0\ ,
\qquad -{1\over3}\leq p\leq1\ ,
\eqno(4.2)
$$
where $\rho$ is again the maximally entangled state in (3.3).
Also in this case one can show that the eigenvalue $\lambda_W(t)$ of
$\rho_W(t)=\Gamma_t[\rho_W]$ corresponding to the eigenvector
$(1,0,0,-1)$ can take negative values, provided the parameter $p$,
that measures the degree of entanglement, is sufficiently close to one.

The discussion becomes particularly transparent when $\alpha=\,0$.
In this case, the eigenvalues of $\rho_W(t)$ become a periodic function
of time; for $\lambda_W(t)$ one then explicitly obtains:
$$
\lambda_W(t)={1\over4}\Bigg\{1+p\Bigg[1-2\bigg(\cos^2 2\Omega t +
{2\omega^2-\Omega^2\over\Omega^2}\sin^2 2\Omega t\bigg)\Bigg]\Bigg\}\ .
\eqno(4.3)
$$
From this expression, one sees that the minimum value of $\lambda_W(t)$
becomes periodically negative provided
$p> (\omega^2-\beta^2)/(\omega^2+3\beta^2)$. This possibility is excluded
in absence of noise: $\beta=\,0$; in this case, the evolution $\gamma_t$
(and hence $\Gamma_t=\gamma_t\otimes\gamma_t$), being unitary,
results automatically positive.

\vskip 1cm

\noindent
{\bf 5. DISCUSSION}
\medskip

The dynamics of a subsystem in weak interaction with an external environment
can be described in terms of linear maps obeying a Markovian master equation.
This result has been rigorously proven in very special cases; it is nevertheless believed to hold in general on the basis of simple physical considerations: 
when all correlations in the environment have died out, non-linearities and memory effects should disappear from the reduced subsytem dynamics.

As shown in Sect.2, the same type of arguments allow deriving a Markovian 
limit for the master equation describing the time evolution of two non-interacting subsystems in contact with the same reservoir; the corresponding dynamical map
$\Gamma_t$ for the compound system turns out to assume a factorized form:
$\Gamma_t=\gamma_t\otimes\gamma_t$.

In the case of two-level systems, $\gamma_t$ usually takes a Bloch-Redfield type form, and therefore it is not in general positive. To avoid inconsistencies, one usually restricts the possible initial states to those for which $\gamma_t$ remains positive (the so-called ``slippage of initial conditions''). This prescription works also in the case of the evolution $\Gamma_t=\gamma_t\otimes\gamma_t$ for two identical subsystems, provided the initial state is in separable form:
$\rho(0)=\sum_i p_i\, \eta^{(1)}_i\otimes\eta^{(2)}_i$,
$p_i\geq0$, $\sum_i p_i=1$, where $\eta^{(1)}_i$ and $\eta^{(2)}_i$
are admissible states for the first and second subsystems, respectively.

On the contrary, as shown in the previous sections, when the initial state $\rho(0)$ is not in factorized form and the degree of entanglement is sufficiently high, the evolved
matrix $\rho(t)=\Gamma_t[\rho(0)]$ fails to be positive for all times.
In keeping with the same attitude adopted for a single subsystem dynamics
$\gamma_t$, to cure this additional inconsistency one can further restrict the domain of applicability of $\gamma_t\otimes\gamma_t$. 
However, this is again a temporary solution: indeed, 
the whole discussion needs be repeated when three or more subsystems in contact with the same bath are considered: clearly,
further restrictions on $\gamma_t$ need to be imposed.

These considerations can not be dismissed as being purely academic;
on the contrary, they seem to have a direct experimental relevance:
as mentioned in the Introduction, couples of systems in an entangled state are in fact actively studied, and 
the ongoing experiments on correlated neutral kaons
constitute a significative example.
From this perspective, the widely used cure of redefining the initial conditions in case of non-positive Markovian dynamics does not appear to be completely satisfactory.

In closing, let us mention that in the few cases for which the Markovian limit of the subdynamics can be obtained in a rigorous way, the resulting evolution map
$\gamma_t$ turns out to be not only positive, 
but also completely positive.[1-4, 11-16] In these cases, the compound map $\Gamma_t=\gamma_t\otimes\gamma_t$ is also completely positive and therefore no inconsistencies can arise, even when $\Gamma_t$ acts on entangled states.[46, 47]

\vskip 1cm

\noindent
{\bf APPENDIX}
\medskip

In Sect.2, we have seen that the master equation that describes the
time evolution of two, non-interacting subsystems in contact with
the same stochastic bath can be written in the following closed form: 
$$
\eqalignno{
&\partial_t\rho(t)=-i\Bigl[H_0^{(1)}+H_0^{(2)}\ ,\ \rho(t)\Bigr]\,-\,
\sum_{A,B=1}^2\sum_{i,j=1}^3C_{ij}^{(AB)}\Bigl[\sigma_i^{(A)},
\Bigl[\sigma_j^{(B)}\ ,\ \rho(t)\Bigr]\Bigr]\ ,
&(A.1)\cr
&C_{ij}^{(AB)}=\sum_{k=1}^3\int_0^\infty{\rm d}s\, W_{ik}^{(AB)}(s)
\,U_{k j}(s)\ , &(A.2)}
$$
after the weak coupling limit and the Markovian 
approximation have been taken into account;
here, $W^{(AB)}_{ij}$ represent the correlation functions in the environment,
while $U_{ij}$ is the orthogonal matrix in (2.5) that takes into account
the rotation of the Pauli matrices generated by the free Hamiltonian.

From the properties (2.3) of the correlation functions, it follows
that the diagonal, on-site coefficients $C^{(AA)}_{ij}$, $A=1,2$,
are real matrices, that can thus be decomposed into symmetric,
${\cal S}^{(A)}_{ij}\equiv \big(C^{(AA)}_{ij} +C^{(AA)}_{ji}\big)/2$,
and antisymmetric,
${\cal A}^{(A)}_{ij}\equiv \big(C^{(AA)}_{ij} -C^{(AA)}_{ji}\big)/2$,
components.
Correspondingly, the second term in $(A.1)$, can be decomposed into
Hamiltonian and purely dissipative pieces, so that the total
master equation can be rewritten as:
$$
\partial_t\rho(t)=-i\bigl[H\, ,\rho(t)\bigr]\,+\,L_D[\rho(t)]\ ,
\eqno(A.3)
$$
where the total Hamiltonian is now given by
$$
H=H_0^{(1)}+H_0^{(2)}+H_D^{(1)}+H_D^{(2)}\ ,\qquad
H_D^{(A)}=\sum_{i,j,k=1}^3 \epsilon_{ijk}\, {\cal A}^{(A)}_{ij}
\, \sigma^{(A)}_k\ ,
\eqno(A.4)
$$
while the dissipative contribution $L_D\equiv L_D^{(1)}+L_D^{(2)}+L_D^{(12)}$
have diagonal and off-diagonal pieces: 
$$
\eqalignno{
&L_D^{(A)}[\rho]=\sum_{ij=1}^3 {\cal S}_{ij}^{(A)}\,
\Big(2\,\sigma_i^{(A)} \rho\, \sigma_j^{(A)} -
\big\{\sigma_i^{(A)}\sigma_j^{(A)}, \rho\big\}\Big)\ ,\qquad A=1,2\ ,&(A.5a)\cr
&L_D^{(12)}=\sum_{ij=1}^3
\Big(C_{ij}^{(12)}+C_{ji}^{(21)}\Big)\Big(
\Big[\sigma_i^{(1)}, \,\rho(t)\, \sigma_j^{(2)}\Big]+
\Big[\sigma_j^{(2)} \rho(t), \,\sigma_i^{(1)}\Big]
\Big)\ .&(A.5b)}
$$
Without additional knowledge on the behaviour of the correlation functions
$W^{(AB)}_{ij}$ of the environment variables, the form of the
evolution equation $(A.3)$--$(A.5)$ can not be further simplified.

On the other hand, with the assumptions (2.6),
$$
W^{(11)}_{33}(t-s)=W^{(22)}_{33}(t-s)=g^2\, e^{-\mu|t-s|}\ ,\qquad
W^{(12)}_{33}(t-s)=f^2\, e^{-\nu|t-s|}\ ,
\eqno(A.6)
$$
and all remaining entries zero, the form of the coefficients
in $(A.2)$ can be explicitly computed, and a more manageable expression
for $(A.3)$ can be obtained. Indeed, after some simple manipulations,
one finds:
$$
\partial_t\rho(t)=\Big(L_0\otimes{\bf 1} + {\bf 1}\otimes L_0\Big)[\rho(t)]+
\Big(L_1\otimes{\bf 1} + {\bf 1}\otimes L_1\Big)[\rho(t)]-
L_2[\rho(t)]\ ,
\eqno(A.7)
$$
where $L_0$ ands $L_1$ are linear operators acting on $2\times2$
density matrices $\eta$
$$
\eqalignno{
&L_0[\eta]=-i\big[H,\eta\big]\ ,\qquad H=\Big({\omega_0\over2}+\beta\Big)\,
\sigma_1\ ,&(A.8a)\cr
&L_1[\eta]=\alpha\big(\sigma_3\,\eta\,\sigma_3-\eta\big)
-\beta\big(\sigma_2\,\eta\,\sigma_3 + \sigma_3\,\eta\,\sigma_2\big)\ , &(A.8b)}
$$
while $L_2$ takes the form:
$$
\eqalign{
L_2[\rho]=\,\gamma&\Big[\big\{\sigma_3\otimes\sigma_3,\, \rho\big\}-
\sigma_3\otimes\sigma_0\,\rho\,\sigma_0\otimes\sigma_3
-\sigma_0\otimes\sigma_3\,\rho\,\sigma_3\otimes\sigma_0\Big]\cr
-\delta&\Big[\big\{\sigma_3\otimes\sigma_2,\, \rho\big\}
+\big\{\sigma_2\otimes\sigma_3,\, \rho\big\}
-\sigma_3\otimes\sigma_0\,\rho\,\sigma_0\otimes\sigma_2\cr
&-\sigma_0\otimes\sigma_2\,\rho\,\sigma_3\otimes\sigma_0
-\sigma_2\otimes\sigma_0\,\rho\,\sigma_0\otimes\sigma_3
-\sigma_0\otimes\sigma_3\,\rho\,\sigma_2\otimes\sigma_0\Big]\ .
}
\eqno(A.8c)
$$
The four constants $\alpha$, $\beta$, $\gamma$, $\delta$,
that measure the relative strength of the various dissipative contributions,
are determined by the parameters appearing in 
the correlation functions $(A.6)$:
$$
\alpha={2 g^2\mu\over\omega_0^2+\mu^2}\ ,\qquad
\beta={g^2\omega_0\over\omega_0^2+\mu^2}\ ,\qquad
\gamma={2 f^2\nu\over\omega_0^2+\nu^2}\ ,\qquad
\delta={f^2\omega_0\over\omega_0^2+\mu^2}\ .
\eqno(A.9)
$$
When the strength and decay constants of the off-diagonal correlations $W_{33}^{(12)}$ are much smaller than the corresponding ones in
$W_{33}^{(AA)}$, $f^2\ll g^2$, $1/\nu\ll 1/\mu$, the dissipative
constants $\gamma$ and $\delta$ can be neglected with respect to
$\alpha$ and $\beta$, and the resulting evolution equation
in $(A.7)$ reduces to that presented in (2.7) and (2.8).

\vfill\eject

\centerline{\bf REFERENCES}
\vskip 1cm

\item{1.} E.B. Davies, {\it Quantum Theory of Open Systems}, (Academic Press,
New York, 1976)
\smallskip
\item{2.} V. Gorini, A. Frigerio, M. Verri, A. Kossakowski and
E.C.G. Surdarshan, Rep. Math. Phys. {\bf 13} (1978) 149 
\smallskip
\item{3.} H. Spohn, Rev. Mod. Phys. {\bf 53} (1980) 569
\smallskip
\item{4.} R. Alicki and K. Lendi, {\it Quantum Dynamical Semigroups and 
Applications}, Lect. Notes Phys. {\bf 286}, (Springer-Verlag, Berlin, 1987)
\smallskip
\item{5.} W.H. Louisell, {\it Quantum Statistical Properties of Radiation},
(Wiley, New York, 1973)
\smallskip
\item{6.} C.W. Gardiner and P. Zoller, {\it Quantum Noise}, II ed., (Springer, Berlin, 2000)
\smallskip
\item{7.} M.O. Scully and M.S. Zubairy, 
{\it Quantum Optics} (Cambridge University Press, Cambridge, 1997)
\smallskip
\item{8.} R.R. Puri, {\it Mathematical Methods of Quantum Optics},
(Springer, Berlin, 2001)
\smallskip
\item{9.} H.-P. Breuer and F. Petruccione, {\it The Theory of Open
Quantum Systems} (Oxford University Press, Oxford, 2002)
\smallskip
\item{10.} R. Dumcke and H. Spohn, Z. Physik {\bf B34} (1979) 419
\smallskip
\item{11.} E.B. Davies, Comm. Math. Phys. {\bf 39} (1974) 91
\smallskip
\item{12.} E.B. Davies, Math. Ann. {\bf 219} (1976) 147
\smallskip
\item{13.} V. Gorini, A. Kossakowski and
E.C.G. Surdarshan, J. Math. Phys. {\bf 17} (1976) 821 
\smallskip
\item{14.} V. Gorini and A. Kossakowski, J. Math. Phys. {\bf 17} (1975) 1298
\smallskip
\item{15.} A. Frigerio and V. Gorini, J. Math. Phys. {\bf 17} (1976) 2123
\smallskip
\item{16.} G. Lindblad, Commun. Math. Phys. {\bf 48} (1976) 119
\smallskip
\item{17.} A.J. van Wonderen and K. Lendi, J. Stat. Phys.
{\bf 80} (1995) 273
\smallskip
\item{18.} A. Royer, Phys. Rev. Lett. {\bf 77} (1996) 3272
\smallskip
\item{19.} D.A. Lidar, Z. Bihary and K.B. Whaley, Chem Phys.
{\bf 268} (2001) 35
\smallskip
\item{20.} S. Gnutzmann and F. Haake, Z. Phys. B {\bf 10} (1996) 263
\smallskip
\item{21.} A. Suarez, R. Silbey and I. Oppenheim, 
J. Chem. Phys. {\bf 97} (1992) 5101
\smallskip
\item{22.} P. Gaspard and M. Nagaoka, J. Chem. Phys. {\bf 111} (1999) 5668
\smallskip
\item{23.} J. Wielkie, J. Chem. Phys. {\bf 114} (2001) 7736
\smallskip
\item{24.} L. Di\'osi, Physica {\bf A199} (1993) 517
\smallskip
\item{25.} {\it The Second Da$\phi$ne Physics Handbook}, L. Maiani, G. Pancheri
and N. Paver, eds., (INFN, Frascati, 1995)
\smallskip
\item{26.} J. Budimir and J.L. Skinner, J. Stat. Phys. {\bf 49} (1987) 1029
\smallskip
\item{27.} B.B. Laird, J. Budimir and J.L. Skinner, J. Chem. Phys. {\bf 94}
(1991) 4391
\smallskip
\item{28.} B.B. Laird and J.L. Skinner, J. Chem. Phys. {\bf 94}
(1991) 4405
\smallskip
\item{29.} J.L. Staudenmann, S.A. Werner, R. Colella and A.W. Overhauser,
Phys. Rev. A {\bf 21} (1980) 1419
\smallskip
\item{30.} S.A. Werner and A.G. Klein, Meth. Exp. Phys. {\bf A23} (1986) 259
\smallskip
\item{31.} V.F. Sears, {\it Neutron Optics}, (Oxford University Press, Oxford,
1989)
\smallskip
\item{32.} {\it Advance in Neutron Optics and Related Research Facilities},
M. Utsuro, S. Kawano, T. Kawai and A. Kawaguchi, eds.,
J. Phys. Soc. Jap. {\bf 65}, Suppl. A, 1996
\smallskip
\item{33.} K.C. Littrell, B.E. Allman and S.A. Werner,
Phys. Rev. A {\bf 56} (1997) 1767
\smallskip
\item{34.} B.E. Allman, H. Kaiser, S.A. Werner, A.G. Wagh, V.C. Rakhecha
and J. Summhammer, Phys. Rev A {\bf 56} (1997) 4420
\smallskip
\item{35.} H. Rauch, S. A. Werner, {\it Neutron Interferometry} (Oxford
University Press, Oxford 2000)
\smallskip
\item{36.} F. Benatti, R. Floreanini and R. Romano, J. Phys. A
{\bf 35} (2002) 4955
\smallskip
\item{37.} E. Collett, {\it Polarized Light}, (Marcel Dekker, New York, 1993)
\smallskip
\item{38.} C. Brosseau, {\it Fundamentals of Polarized Light},
(Wiley, New York, 1998)
\smallskip
\item{39.} F. Benatti and R. Floreanini,
Banach Center Publications, {\bf 43} (1998) 71
\smallskip
\item{40.} F. Benatti and R. Floreanini, Nucl. Phys. {\bf B511} (1998) 550
\smallskip
\item{41.} F. Benatti and R. Floreanini, Ann. of Phys. {\bf 273} (1999) 58
\smallskip
\item{42.} F. Benatti, R. Floreanini and R. Romano, 
Nucl. Phys. {\bf B602} (2001) 541
\smallskip
\item{43.} T. Yu, L. Diosi, N. Gisin and W.T. Strunz, Phys. Rev. A
{\bf 60} (1999) 91
\smallskip
\item{44.} T. Yu, L. Diosi, N. Gisin and W.T. Strunz,
Phys. Lett. {\bf A265} (2000) 331
\smallskip
\item{45.} R.F. Werner, Phys. Rev. A {\bf 40} (1989) 4277
\smallskip
\item{46.} F. Benatti and R. Floreanini, Chaos, Solitons and Fractals 
{\bf 12} (2001) 2631
\smallskip
\item{47.} F. Benatti, R. Floreanini and R. Romano, J. Phys. A
{\bf 35} (2002) L551

\bye